\documentclass[%
 reprint,
 amsmath,amssymb,
 aps,
]{revtex4-2}
\usepackage{overpic}
\usepackage{graphicx}% Include figure files
\usepackage{dcolumn}% Align table columns on decimal point
\usepackage{bm}% bold math
\begin{document}

\preprint{APS/123-QED}

\title{High quality beam produced by tightly focused laser driven wakefield accelerators}

\author{Jia Wang}
\author{Ming Zeng}%
 \email{zengming@ihep.ac.cn}
\author{Dazhang Li}
 \email{lidz@ihep.ac.cn}
\author{Xiaoning Wang}
\author{Jie Gao}
\affiliation{%
 Institute of High Energy Physics, Chinese Academy of Sciences, Beijing 100049, China
}%
\affiliation{University of Chinese Academy of Sciences, Beijing 100049, China
}%

\date{\today}

\begin{abstract}
We propose to use tightly focused lasers to generate high quality electron beams in laser wakefield accelerators. In this scheme, the expansion of the laser beam after the focal position enlarges the size of wakefield bubble, which reduces the effective phase velocity of the wake and triggers injection of plasma electrons. This scheme injects a relatively long beam with high charge. The energy spread of the injected beam can be minimized if an optimal acceleration distance is chosen so that the beam chirp is suppressed. Particle-in-cell simulations indicate that electron beams with the charge in the order of nanocoulomb, the energy spread of $\sim 1\%$, and the normalized emittance of $\rm \sim 0.1\ mm\cdot mrad$ can be generated in uniform plasma using $\sim 100\ \rm TW$ laser pulses. An empirical formula is also given for predicting the beam charge. This injection scheme, with a very simple setup, paves the way towards practical high-quality laser wakefield accelerators for table-top electron and radiation sources.
\end{abstract}

\maketitle

%\tableofcontents
\section{\label{sec:level1}INTRODUCTION}

Laser wakefield accelerators (LWFAs), proposed in 1979, have attracted wide attention due to their 3 to 4 orders of magnitude higher acceleration gradient than that of the conventional radio-frequency (RF) accelerators~\cite{TTajimaPRL1979}. Recently, $8\ \rm GeV$ electron beams have been obtained by an LWFA within $20\ \rm cm$ acceleration distance~\cite{AJGonsalvesPRL2019}. Successive 24 h stable operation of LWFA and free-electron-laser based on LWFA have shown the bright prospect of this acceleration mechanism~\cite{MaiPRL2020, WTWangNature2021}. Although the output beam quality of the LWFA has been largely improved~\cite{KeLPRL2021}, there is still a certain distance for competing with the RF accelerators and satisfying the requirements of real applications. Obtaining high-quality stable electron bunches has always been a consistent goal of this research field.

In an LWFA, the ponderomotive force of a high intensity laser pulse off-axially expels the background electrons of an underdense plasma. Afterwards they are pulled back by the nearly immobile ions and form an oscillating plasma wake wave following the drive laser. Under highly nonlinear conditions, the plasma electrons can be completely evacuated to form an ion column where strong acceleration field is formed. Bunched electrons can gain energies of several GeV within a few centimeters if they are placed at appropriate accelerating and focusing phases by a certain injection scheme. The injection scheme directly influences the quality of the electron bunch. In the past decades, many injection schemes have been proposed. The self-injection scheme uses the wave-breaking effect when the drive pulse reaches a certain threshold~\cite{LeeNP2006,WangNC2013,LeePRL2014,KalPRL2009,KalPPCF2011,KalPOP2011}. The optical injection schemes pre-accelerate the electrons by the beat wave of counter-propagating laser pulses or the ponderomotive force of assistant laser pulses so that the injection threshold of the wakefield potential is reduced~\cite{SchPRE1999,GolPRL2018, UmsPRL1996, ZengNJP2020}. The ionization injection scheme uses the electric field of the driver to ionize inner-shell electrons of dopant gas atoms inside the wake to partially avoid the deceleration phase in the first half period of the wakefield, so that these electrons are more likely to gain enough forward velocity for injection~\cite{ChenJAP2006,PakPRL2010,WangJPPCF2022,ZengMPOP2014}. The density gradient injection scheme reduces the phase velocity of the wakefield by introducing a density decreasing region~\cite{MasPPCF2017,KeLPRL2021}. And the coaxial laser interference injection scheme reduces the phase velocity of the wakefield by the evolution of interference rings created by the tightly focused trigger laser which is coaxially propagating with the drive laser~\cite{WangJMRE2022}.

In this paper, we propose a new injection scheme that utilizes the evolution of a tightly focused drive laser to trigger the injection of background plasma electrons in LWFAs. Unlike typical self-injection due to the bubble evolution~\cite{KalPRL2009}, our injection process is more predictable because the injection is confined within a few Rayleigh range of the tightly focused drive laser, where the main reason of bubble evolution is the laser defocusing after the focal point, instead of highly nonlinear laser-plasma interaction. Such laser defocusing leads to smooth expansion of the wakefield bubble and to the injection of electron beams with the charge in the order of nanocoulomb (nC), normalized emittance in the order of $0.1\ \rm mm \cdot mrad$, and slice energy spread in the order of $0.1\%$. At the optimal acceleration distances, the energy spread of the whole beam is minimized to the order of $1\%$ due to the self-dechirping effect. Our scheme simultaneously produces high charge and small energy spread electron beams with a very basic setup of LWFA, thus has potential of wide applications.

\section{\label{sec:level2}A Phenomenological theory}

Previous studies have shown that electron beams with charge on the order of $\rm nC$ can be generated in underdense plasma with density of $n_p\sim 10^{19}$ to $10^{20}\ \rm cm^{-3}$ by tightly focused laser pulses with waist radius $w_0$ close to the laser wavelength $\lambda$ and power of $P\approx 100\ \rm TW$, but the quality of the beams is relatively unsatisfactory~\cite{XuHPOP2005,ZhiPRE2008}. The electron thermalization in underdense plasma is responsible for the broad energy spectrum, and the small laser spot is also disadvantageous to the energy chirp reduction of electron beam~\cite{ZhiPRE2008}.

Here we explore the injection with plasma density $n_p\approx 10^{18}\rm cm^{-3}$ and laser waist radius $w_0$ between $2$ and $10\ \mathrm{\mu m}$ for laser wavelength $800\ \rm nm$. We consider the region with $da/dz<0$, where $a=eA/m_ec^2$ is the normalized amplitude of the vector potential of the laser pulse (or the laser strength parameter), $e$ is elementary charge, $A$ is the amplitude of laser vector potential, $m_e$ is the rest mass of electrons, $c$ is the speed of light. For a tightly focused laser pulse with power $\sim 100\ \rm TW$, the ponderomotive force is strong enough to evacuate the background plasma electrons and create the electron-free bubble, so that the laser defocuses within a short range as if it is in vacuum~\cite{ZengMPOP2020}.

The transverse component of ponderomotive force of a Gaussian laser pulse can be written as~\cite{MP1997}
\begin{equation}
  \begin{aligned}
    F_{pr}&=-\frac{m_ec^2a^2}{4\left<\gamma\right>}\frac{\partial}{\partial r}\exp\left(-2\frac{r^2}{w^2}\right) \\
    &=\frac{m_ec^2a_0^2w_0^2r}{w^4\left<\gamma\right>}\exp\left(-2\frac{r^2}{w^2}\right), \label{eq:pon}
  \end{aligned}
\end{equation}
where $a_0$ is the value of $a$ at focus in vacuum, $w_0$ is the laser waist radius in vacuum, $r=\sqrt{x^2+y^2}$ is the transverse position, $\left<\gamma\right>$ is the Lorentz factor averaged in one laser cycle, $w$ is the laser radius which is a function of the longitudinal coordinate $z$. Due to the effect of passive plasma lens for laser, the laser is focused more tightly in plasma than in vacuum~\cite{PalPOP2015, ZengMPOP2020}, thus the actual evolution of the radius in the first laser envelope oscillation period can be approximately written as
\begin{equation}
    w\left(z\right)=w_0\Gamma\sqrt{1+(z-z_{fe})^2/z_{Re}^{2}}, \label{eq:wz}
\end{equation}
where $z_{fe}$ is the effective focal position in plasma, $z_{Re}=\pi w_0^2\Gamma^2/\lambda$ is the effective Rayleigh length. Neglect the energy loss of the laser, $\Gamma$ can be written as 
\begin{equation}
    \Gamma=\frac{a_0}{a_{0e}}=\frac{w_{0e}}{w_0}, \label{eq:f}
\end{equation}
where $a_{0e}$ and $w_{0e}$ are the effective peak strength parameter and the effective waist radius at $z_{fe}$. We know that $\Gamma<1$ if the focal position is inside the plasma, and $\Gamma=1$ if it is outside. The focusing force in the bubble is~\cite{LuW2006}
\begin{equation}
    F=-\kappa^2r,\label{eq:Ff}
\end{equation}
where $\kappa^2=\omega_p^2m_e/\alpha$, and $\omega_p$ is the plasma frequency. Generally $\alpha\geqslant2$, and $\alpha=2$ corresponds to the case of electron-free ion cavity. Balancing the ponderomotive force Eq.~(\ref{eq:pon}) and the focusing force Eq.~(\ref{eq:Ff}), we can approximately get the bubble radius 
\begin{equation}
    r_{b}=\sqrt{\frac{w^2}{2}\ln\left(\frac{\Omega w_0^2a_0^2}{w^4k_p^2}\right)},\label{eq:rbl}
\end{equation}
where $\Omega=\left.\alpha/\left<\gamma\right>\right|_{r=r_b}$, and $k_p=\omega_p/c$ is the wavenumber of the plasma wave. For simplicity, we assume $\Omega$ is a constant. By taking derivative of Eq.~(\ref{eq:rbl}), we know $dr_{b}/dz>0$ if $\ln\left(\Omega w_0^2 a_0^2\right)-2>4\ln w$. In other words,
\begin{equation}
    \left.w\right|_{z=z_{ie}}=\exp\left(-\frac{1}{2}\right)\left(\frac{\Omega w_0^2a_0^2}{k_p^2}\right)^{\frac{1}{4}},\label{eq:zie}
\end{equation}
where $z_{ie}$ is the ending point of the bubble expansion.
The transverse expansion of the bubble leads to longitudinal expansion also, which decreases the phase velocity of the wakefield and triggers the injection of the electrons into the bubble. By solving Eqs.~(\ref{eq:wz}) and (\ref{eq:zie}), an optimistic estimation of injection length is obtained 
\begin{equation}
  \begin{aligned}
    L_{inj}&=z_{ie}-z_{fe} \\
    &=z_{Re}\sqrt{\exp\left(-1\right)\frac{\sqrt{\Omega} a_0}{\Gamma^2k_pw_0}-1}.\label{eq:Linj}
  \end{aligned}
\end{equation}

\begin{figure}
    \centering
    \begin{overpic}[width=0.48\textwidth]{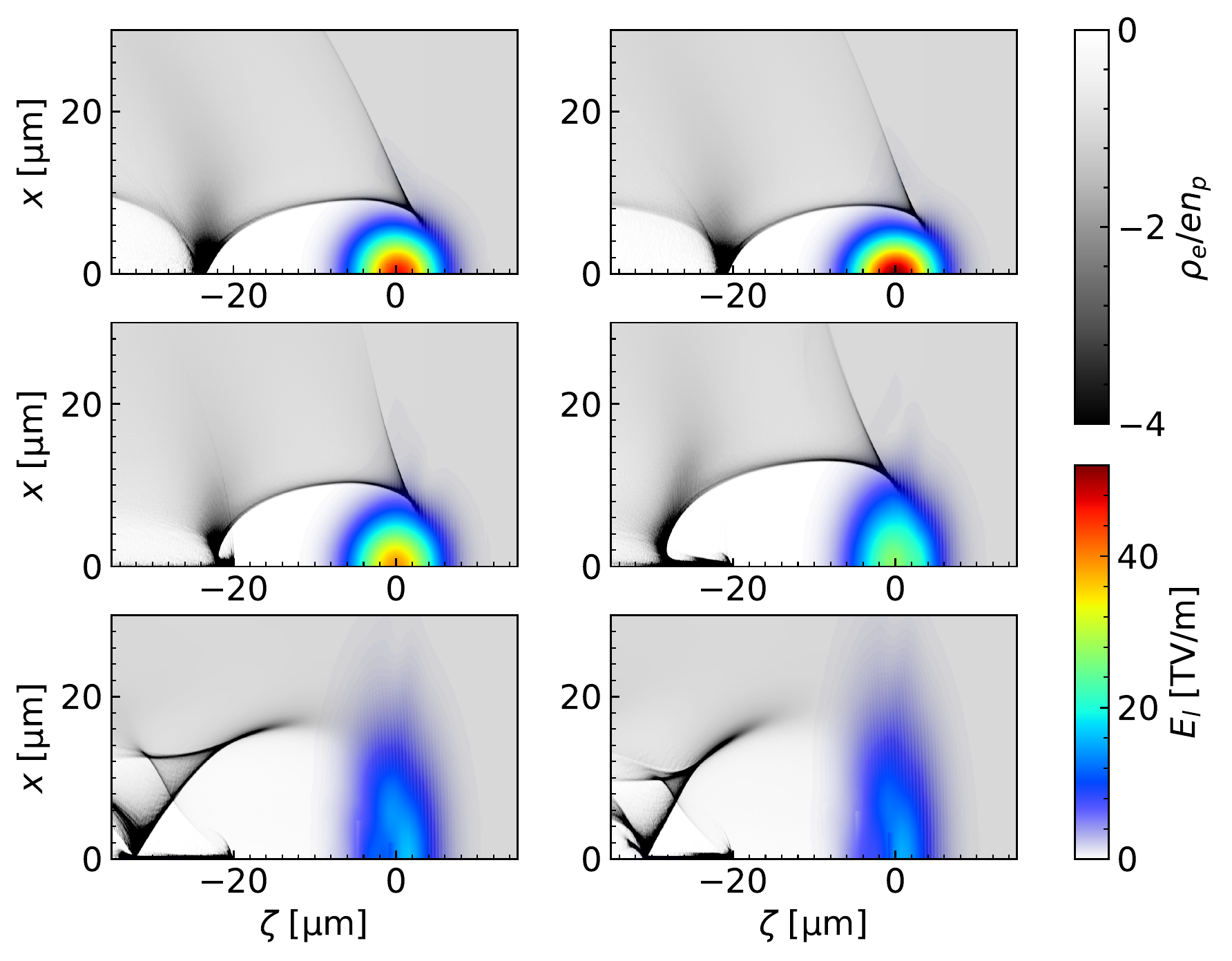}
       \put(11,72){$z_l=160\ \rm \mu m$}
       \put(52,72){$z_l=192\ \rm \mu m$}
       \put(11,48){$z_l=269\ \rm \mu m$}
       \put(52,48){$z_l=329\ \rm \mu m$}
       \put(11,24){$z_l=479\ \rm \mu m$}
       \put(52,24){$z_l=532\ \rm \mu m$}
    \end{overpic}
    \caption{\label{fig:6snap} The snapshots of the injection process in the plasma wakefield driven by a tightly focused laser. $\rho_e$ is the electron density, $n_p$ is the undisturbed plasma density, $E_l$ is the profile of the electric field of the laser, and $\zeta=z-ct$ is the comoving coordinate. The laser center positions $z_l$ are written for each of the subplots.
}
\end{figure}

\begin{figure}
    \centering
    \begin{overpic}[width=0.48\textwidth]{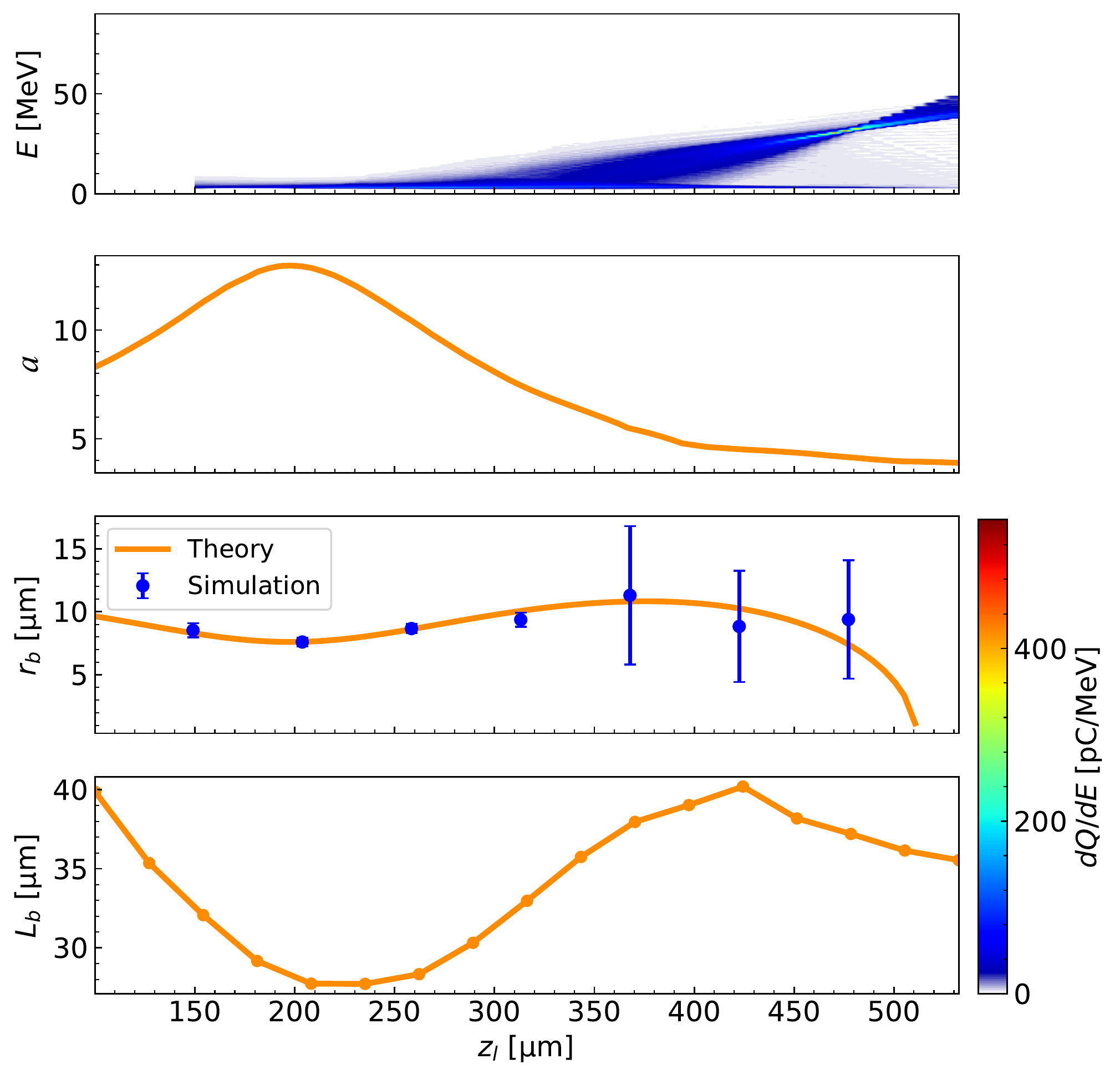}
        \put(11,81){(a)}
        \put(11,56){(b)}
        \put(11,33){(c)}
        \put(11,10){(d)}
    \end{overpic}
    \caption{\label{fig:Es} The evolution of (a) the energy spectrum of the trapped electrons, (b) the laser strength parameter $a$, (c) the bubble radius $r_{b}$, and (d) the bubble length $L_b$ vs. the position of the laser center $z_l$. The errorbar in (c) shows the range of the bubble sheath.
}
\end{figure}

\begin{figure}
    \centering
    \begin{overpic}[width=0.48\textwidth]{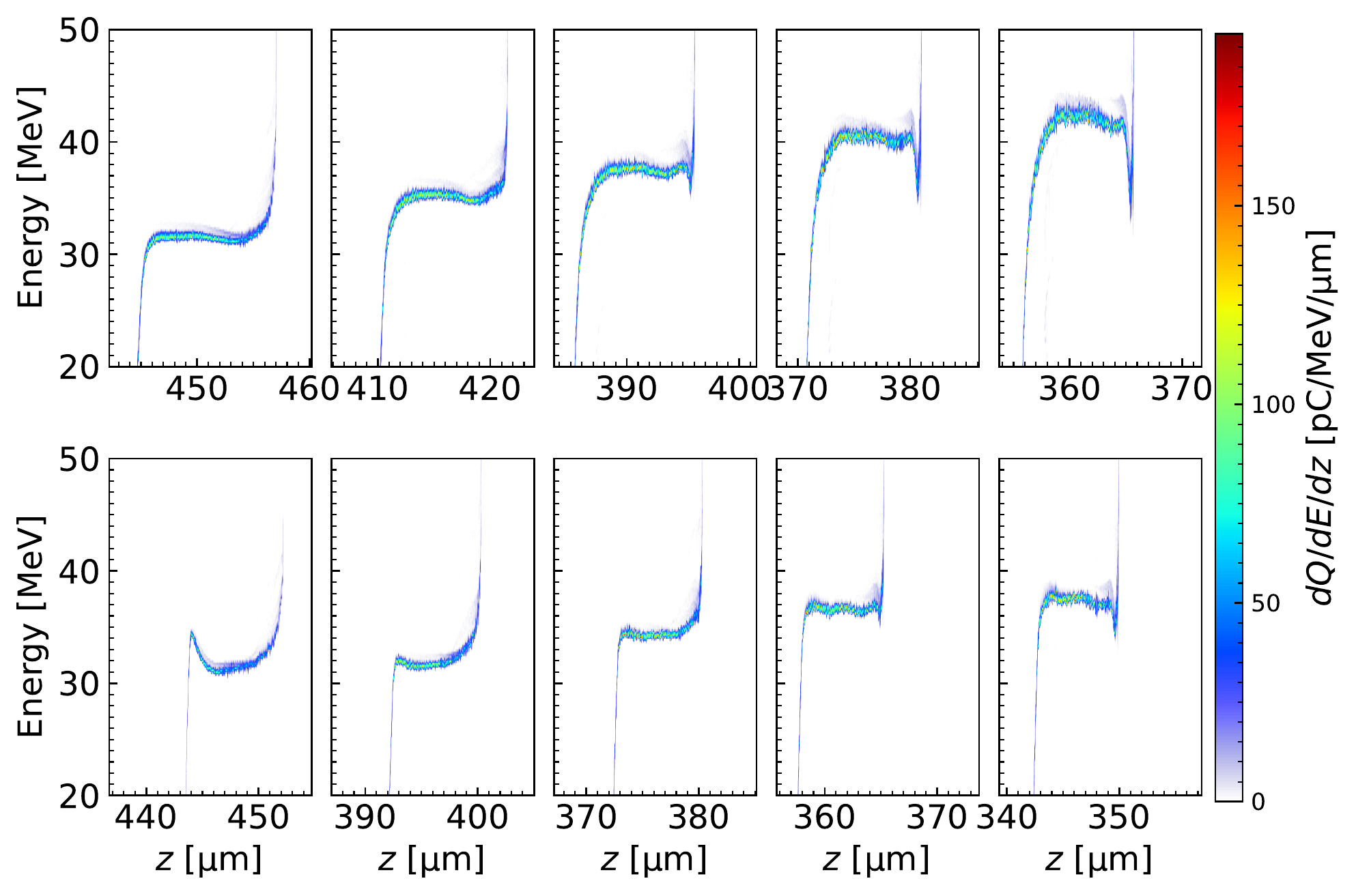}
        \put(8,60){(a-1)}
        \put(8.5,41){$\rm 560\ pC$}
        \put(24.5,60){(a-2)}
        \put(25,41){$\rm 720\ pC$}
        \put(41,60){(a-3)}
        \put(41.5,41){$\rm 810\ pC$}
        \put(57.5,60){(a-4)}
        \put(58,41){$\rm 880\ pC$}
        \put(74,60){(a-5)}
        \put(74.5,41){$\rm 940\ pC$}
        \put(8,28){(b-1)}
        \put(8.5,9){$\rm 275\ pC$}
        \put(24.5,28){(b-2)}
        \put(25,9){$\rm 360\ pC$}
        \put(41,28){(b-3)}
        \put(41.5,9){$\rm 420\ pC$}
        \put(57.5,28){(b-4)}
        \put(58,9){$\rm 475\ pC$}
        \put(74.5,28){(b-5)}
        \put(75,9){$\rm 520\ pC$}
    \end{overpic}
    \caption{\label{fig:p1} The electron beam phase spaces at optimal acceleration lengths, which minimize the energy spreads, for different cases with laser power (a) $P=120\ \rm TW$ ($a_0=12$, $w_0=5\ \rm\mu m$) and (b) $P=60\ \rm TW$ ($a_0=8.45$, $w_0=5\ \rm\mu m$). The plasma densities $n_p$ for (x-1 to 5) are $2\times 10^{18}\rm cm^{-3}$, $3\times 10^{18}\rm cm^{-3}$, $4\times 10^{18}\rm cm^{-3}$, $5\times 10^{18}\rm cm^{-3}$ and $6\times 10^{18}\rm cm^{-3}$, respectively, where x stands for a or b.
}
\end{figure}

\begin{figure}
    \centering
    \begin{overpic}[width=0.48\textwidth]{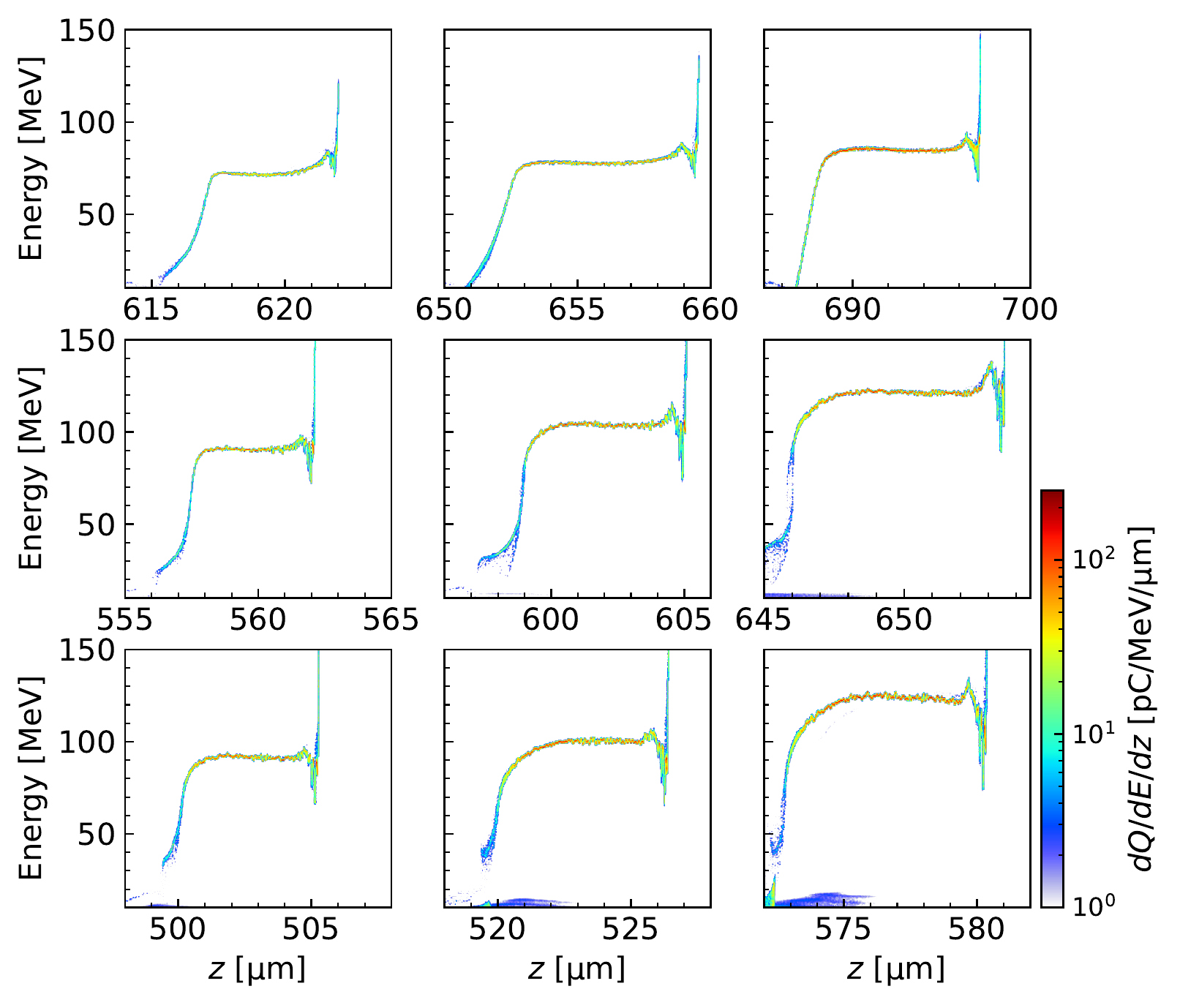}
       \put(11.5,79){(a-1)}
       %\put(13,67){$\rm 0.35\ nC$}
       %\put(11.5,63){$\delta_e=0.93\%$}
       \put(38.5,79){(a-2)}
       %\put(40,67){$\rm 0.58\ nC$}
       %\put(38.5,63){$\delta_e=1.03\%$}
       \put(65.5,79){(a-3)}
       %\put(67,67){$\rm 0.85\ nC$}
       %\put(65.5,63){$\delta_e=0.94\%$}
       \put(11.5,52.7){(b-1)}
       %\put(13,40){$\rm 0.46\ nC$}
       %\put(11.5,36){$\delta_e=0.84\%$}
       \put(38.5,52.7){(b-2)}
       %\put(40,40){$\rm 0.73\ nC$}
       %\put(38.5,36){$\delta_e=1.2\%$}
       \put(65.5,52.7){(b-3)}
       %\put(67,40){$\rm 1\ nC$}
       %\put(65.5,36){$\delta_e=0.86\%$}
       \put(11.5,26.4){(c-1)}
       %\put(13,14.4){$\rm 0.6\ nC$}
       %\put(11.5,10.4){$\delta_e=1.32\%$}
       \put(38.5,26.4){(c-2)}
       %\put(40,14.4){$\rm 0.9\ nC$}
       %\put(38.5,10.4){$\delta_e=1.27\%$}
       \put(65.5,26.4){(c-3)}
       %\put(67,14.4){$\rm 1.2\ nC$}
       %\put(65.5,10.4){$\delta_e=1.3\%$}
    \end{overpic}
    \caption{\label{fig:Q3} The electron beam phase spaces at optimal acceleration lengths, which minimize the energy spreads, for fixed $w_0=7\ \rm \mu m$ but different $a_0$ and $n_p$. For (a), (b) and (c), $n_p$ equals to $2\times 10^{18}\ \rm cm^{-3}$, $4\times 10^{18}\ \rm cm^{-3}$ and $6\times 10^{18}\ \rm cm^{-3}$, respectively. And for (x-1, 2, 3), $a_0$ equals to 8, 10 and 12, respectively, where x stands for a, b or c.
}
\end{figure}

\begin{figure}
    \centering
    \begin{overpic}[width=0.48\textwidth]{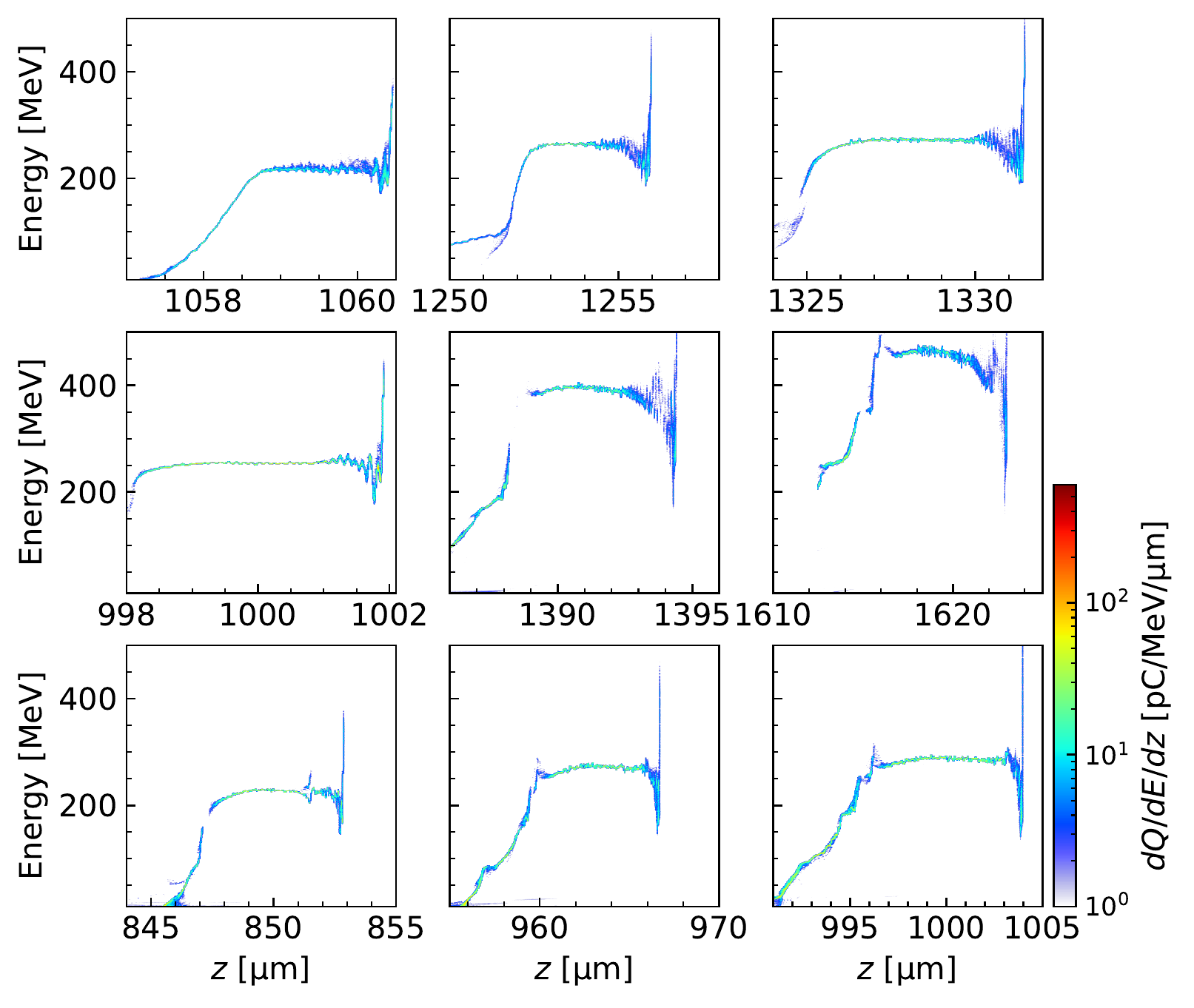}
       \put(11.5,79){(a-1)}
       \put(38.5,79){(a-2)}
       %\put(40,67){$\rm 0.62\ nC$}
       %\put(38.5,63){$\delta_e=1.39\%$}
       \put(65.5,79){(a-3)}
       %\put(67,67){$\rm 1\ nC$}
       %\put(65.5,63){$\delta_e=1.03\%$}
       \put(11.5,52.7){(b-1)}
       %\put(13,40){$\rm 0.63\ nC$}
       %\put(11.5,36){$\delta_e=0.94\%$}
       \put(38.5,52.7){(b-2)}
       \put(65.5,52.7){(b-3)}
       \put(11.5,26.4){(c-1)}
       \put(38.5,26.4){(c-2)}
       \put(65.5,26.4){(c-3)}
    \end{overpic}
    \caption{\label{fig:Q4} The electron beam phase spaces at optimal acceleration lengths, which minimize the energy spreads, for fixed $w_0=9\ \rm \mu m$ but different $a_0$ and $n_p$. For (a), (b) and (c), $n_p$ equals to $2\times 10^{18}\ \rm cm^{-3}$, $4\times 10^{18}\ \rm cm^{-3}$ and $6\times 10^{18}\ \rm cm^{-3}$, respectively. And for (x-1, 2, 3), $a_0$ equals to 8, 10 and 12, respectively, where x stands for a, b or c.
}
\end{figure}

\begin{figure}
    \centering
    \begin{overpic}[width=0.48\textwidth]{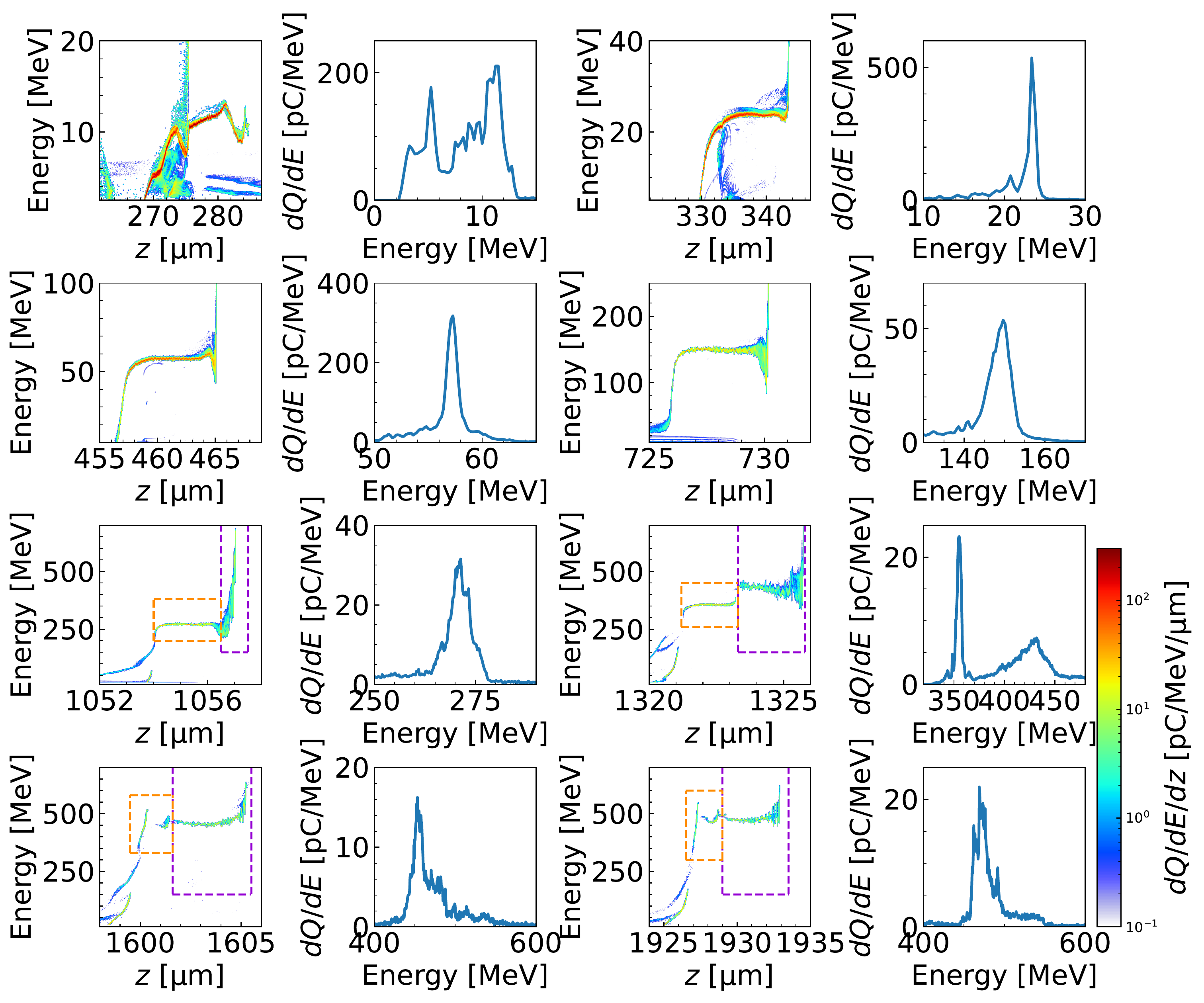}
       \put(9,76){(a-1)}
       %\put(11.5,59){$w_0=2\ \mu m$}
       \put(32,76){(a-2)}
       \put(55,76){(b-1)}
       %\put(56.8,59){$w_0=4\ \mu m$}
       \put(78,76){(b-2)}
       \put(9,56){(c-1)}
       %\put(11.5,43.5){$w_0=6\ \mu m$}
       \put(32,56){(c-2)}
       \put(55,56){(d-1)}
       %\put(56.8,43.5){$w_0=8\ \mu m$}
       \put(78,56){(d-2)}
       \put(9,36){(e-1)}
       %\put(11.5,28){$w_0=10\ \mu m$}
       \put(32,36){(e-2)}
       \put(55,36){(f-1)}
       %\put(56.8,28){$w_0=12\ \mu m$}
       \put(78,36){(f-2)}
       \put(9,16){(g-1)}
       %\put(11.5,13){$w_0=15\ \mu m$}
       \put(32,16){(g-2)}
       \put(55,16){(h-1)}
       %\put(56.8,13){$w_0=20\ \mu m$}
       \put(78,16){(h-2)}
    \end{overpic}
    \caption{\label{fig:Q5} The phase spaces and energy spectra of the injected electron beams for fixed laser peak power $\rm 120\ TW$ and plasma density $n_p=4\times 10^{18}\ \rm cm^{-3}$, while $a_0$ and $w_0$ from (a) to (h) are (30, $2\ \rm \mu m$), (15, $4\ \rm \mu m$), (10, $6\ \rm \mu m$), (7.5, $8\ \rm \mu m$), (6, $10\ \rm \mu m$), (5, $12\ \rm \mu m$), (4, $15\ \rm \mu m$) and (3, $20\ \rm \mu m$), respectively. For $4\ \mathrm{\mu m}\leq w_0\leq8\ \mathrm{\mu m}$ cases, all of the electrons have the injection positions between $z_{fe}$ and $z_{ie}$. For larger $w_0$, however, other injection positions can be majorities. The electrons in the violet dashed rectangular boxes have the injection positions before $z_{fe}$, and the ones in the orange dashed rectangular boxes have the injection positions between $z_{fe}$ and $z_{ie}$.
}
\end{figure}

\section{\label{sec:level3}PIC simulations}
%\subsection{\label{sec:level2-B}PIC simulations}
A series of quasi-cylindrical particle-in-cell (PIC) simulations using the code WarpX with the Pseudo-Spectral Analytical Time Domain (PSATD) solver have been carried out, where two azimuthal modes are used~\cite{JLVayPOP2021, RemCPC2016}. An example is illustrated in Figs.~\ref{fig:6snap} and \ref{fig:Es}. The simulation box has the size of ($\rm 50\ \mu m$, $\rm 50\ \mu m$) and the cell number of (3200, 512) in $z$ and $r$ directions, respectively. The number of particles per cell along $\left(z, r, \theta\right)$ directions are $\left(2, 2, 4\right)$. The plasma density profile has a liner upramp from $z=0$ to $z=\rm100\ \mu m$, followed by a flattop with the density $n_p= 2\times10^{18}\rm cm^{-3}$. The laser pulse has $a_0=12$, $w_0=5\rm \ \mu m$, $\lambda=800\ \rm nm$ (the peak power is about 120 TW), $z_f=200\rm \ \mu m$ and the pulse duration in full-width-half-maximum $\tau=20\ \rm fs$. After the effective laser focal point ($z_{fe}\approx z_f$ for this case), $a$ decreases, $w$ increases, and $r_{b}$ increases according to Eq.~(\ref{eq:rbl}) which is in agreement with the simulation result as shown in Fig.~\ref{fig:Es}(c). Later $r_b$ decreases after the expansion ending point $z_{ie}$ which can be obtained by solving Eq.~(\ref{eq:zie}). From Fig.~\ref{fig:Es}(c) and (d) one can see that the longitudinal size of the bubble $L_b$ is strongly correlated with its transverse size $r_b$. The increasing starting/ending points of $L_b$ are almost the same as the increasing starting/ending points of $r_b$. Thus, the injection length, which is no longer than one increasing period of $L_b$, can be estimated by Eq.~(\ref{eq:Linj}). At the tail of the beam, where the energy is initially lower, the higher acceleration gradient compensates the energy chirp of the whole beam. A high quality electron beam with charge of $Q\approx \rm 560\ pC$, root-mean-squared (RMS) energy spread of $0.95\%$ (obtained by Gaussian fit) and normalized emittance of $\epsilon_x=0.34\ \rm mm\cdot mrad$ and $\epsilon_y=0.05\ \rm mm\cdot mrad$ is generated at $z_l=480\ \rm \mu m$ as shown in Fig.~\ref{fig:Es}(a) and Fig.~\ref{fig:p1}(a-1), where $z_l$ is the position of the laser pulse. 
%And the energy is close to the Eqs.~(\ref{eq:EEf}) of $E_f= 27.5\ \rm MeV$.

The phase spaces of the trapped electron beams at different plasma densities and laser powers are shown in Fig.~\ref{fig:p1}. The beam charge is positively correlated with the laser intensity and plasma density. At different optimal acceleration lengths, which are in the order of hundred micrometers in these cases, the minimal energy spreads of the bunches are achieved, which are close to the slice energy spreads.
In addition to the high-quality injection within the first bubble expansion period $z_{fe}<z<z_{ie}$, there can be low-quality injections at later bubble evolution periods due to laser-plasma interaction. These later injected beams can be separated from the high-quality injection due to their energy differences.

The relative low beam energies for $w_0=5\ \rm \mu m$ cases are due to the short accelerating lengths. With an increasing $w_0$, the beam energy and optimal length increase as shown in Figs.~\ref{fig:Q3} and \ref{fig:Q4}. The electron beam can have a monoenergetic peak in the spectrum with the central energy in the order of 10 to 100 MeV and the charge in the order of nanocoulomb. For example, the laser with the peak power of $P= 237\ \rm TW$ generates an electron beam with the energy of $ 90\ \rm MeV$, the charge of $\rm 0.85\ nC$ and the energy spread of $0.94\%$ as shown in Fig.~\ref{fig:Q3}(a-3). The laser with $P= 174\ \rm TW$ generates electron beam with the energy of $250\ \rm MeV$, the charge of $\rm 0.63\ nC$ and the energy spread of $0.94\%$ as shown in Fig.~\ref{fig:Q4}(b-1). The wave breaking leads to the density spike at the head of the electron beam, which has an attosecond duration and ultrahigh current ($\gtrsim\rm 100\ kA$). The electrons injected after $z_{ie}$ form the long tail of the beam as shown in Fig.~\ref{fig:Q4}(b-2)(b-3) and (c-1)-(c-3), which broaden the energy spectra. 

The spectra and phase spaces of electron beams from different cases with a fixed laser power of $\rm 120\ TW$ are shown in Fig.~\ref{fig:Q5}. 
The case with $w_0=2\ \rm \mu m$ is more similar to the ``cube of the pulse wavelength'' injection mechanism~\cite{ZhiPRE2008}. For $w_0\geq 10\ \mathrm{\mu m}$, the injections come from both $z<z_{fe}$ and $z_{fe}<z<z_{ie}$ positions, and the portion of injection from $z<z_{fe}$ becomes more significant when $w_0$ increases further. The moderately tightly focused cases, with $4\ \mathrm{\mu m}\leq w_0\leq8\ \mathrm{\mu m}$, have the injection mechanism proposed in the present work. They have the injection positions dominantly between $z_{fe}$ and $z_{ie}$, and each of them has one mono-energetic peak in the spectrum.

\section{\label{sec:level4}charge scaling}
The amount of trapped charge is a key parameter for plasma based accelerators. For a static bubble under matched conditions, the number of electrons trapped in the accelerating phase can be estimated by equating the electromagnetic field energy in the ions cavity with the energy absorbed by the particles~\cite{LuW2007}.
The injection scheme in this work is different, because the bubble is expanding. We look for an empirical formula for the injected beam in this section. 

According to the simulation observations and previous experimental facts~\cite{KusPRL2018,FroPRL2009}, we can reasonably assume that the amount of injected charge is linearly correlated with the production of the injection length $L_{inj}$, the plasma density $n_p$ and the laser strength parameter at focus $a_0$, written as
\begin{equation}
    Q=C\times L_{inj}n_pa_0,\label{eq:Q}
\end{equation}
where $C$ is an empirical constants. We notice that $\Gamma$ and $\Omega$ also have to be determined to calculate $Q$. To simplify the model, we assume $\Omega=2$ and do not consider the dependencies of $\Gamma$ on the laser focal position $z_f$ and the laser wavelength $\lambda$ (fixed at $800\ \rm nm$). Then $\Gamma$ is a functions of $n_p$, $a_0$ and $w_0$ which can be approximately written in the following empirical form within a certain parameter space (see Supplementary Material)
\begin{equation}
%\Gamma=\frac{n_p}{C_n}+\frac{a_0}{C_a}+\frac{w_0}{C_w} + \Gamma_0,\label{eq:a1}
\Gamma\approx-\frac{n_p\left[10^{18}\ \rm cm^{-3}\right]}{20.16}+\frac{a_0}{100}-\frac{w_0\left[\rm\mu m\right]}{46.42}+1.029.\label{eq:ff}
\end{equation}

\begin{figure}
    \centering
    \begin{overpic}[width=0.48\textwidth]{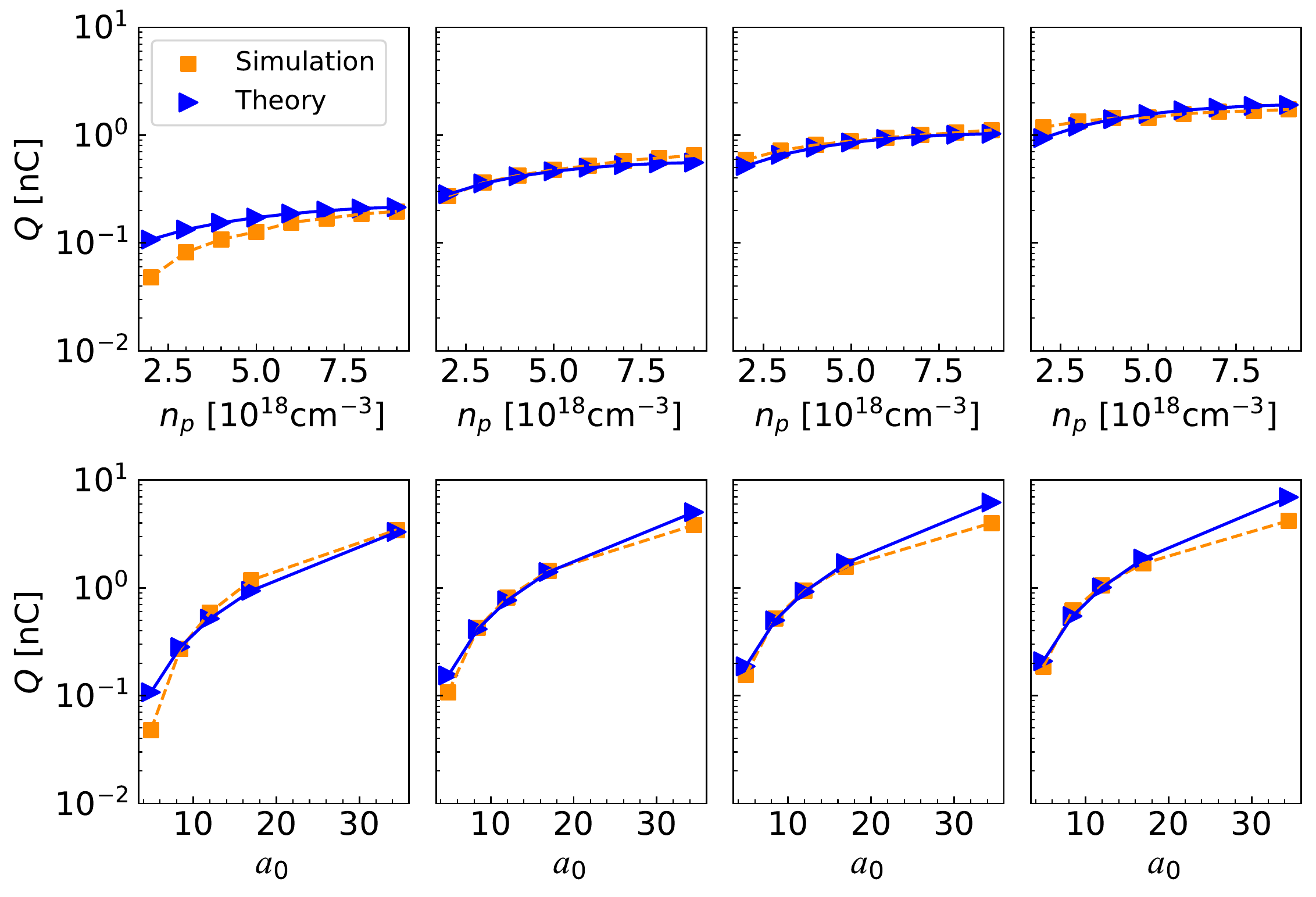}
       \put(12,43.2){(a-1)}
       \put(34.6,43.2){(a-2)}
       \put(57.2,43.2){(a-3)}
       \put(79.8,43.2){(a-4)}
       \put(12,8.5){(b-1)}
       \put(34.6,8.5){(b-2)}
       \put(57.2,8.5){(b-3)}
       \put(79.8,8.5){(b-4)}
    \end{overpic}
    \caption{\label{fig:Q1} The injected charge vs.\ (a) the plasma density $n_p$ and (b) the laser strength parameter at focus $a_0$ with a fixed laser waist radius $w_0=\rm 5\ \mu m$. For (a-1 to 4), $a_0$ and the corresponding powers are (4.9, $\rm 20\ TW$), (8.45, $\rm 60\ TW$), (12, $\rm 120\ TW$) and (16.97, $\rm 240\ TW$), respectively. And for (b-1 to 4), $n_p$ equals to $2\times 10^{18}\rm cm^{-3}$, $4\times 10^{18}\rm cm^{-3}$, $6\times 10^{18}\rm cm^{-3}$ and $8\times 10^{18}\rm cm^{-3}$, respectively.
}
\end{figure}

\begin{figure}
    \centering
    \begin{overpic}[width=0.48\textwidth]{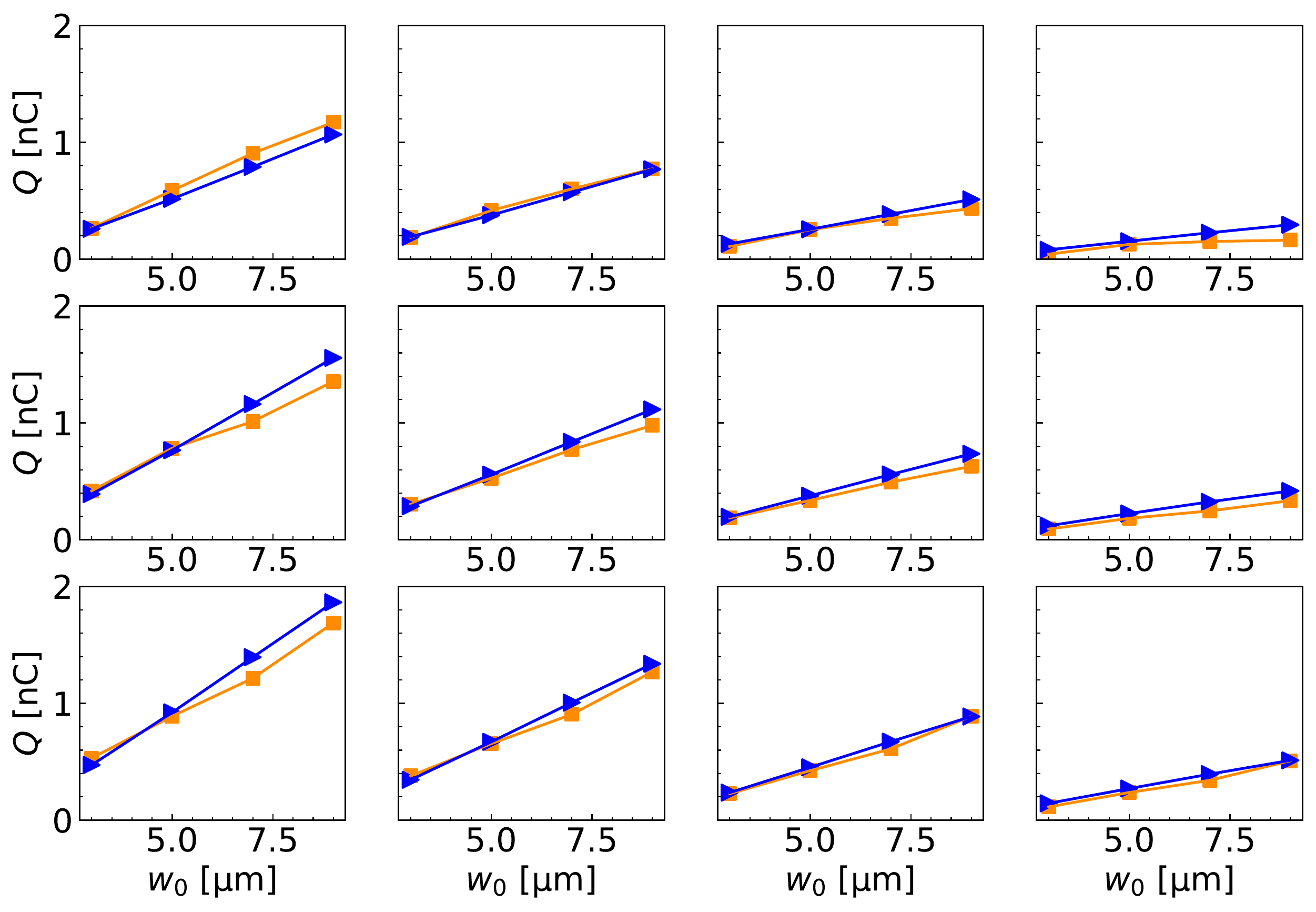}
       \put(6.4,63.5){(a-1)}
       \put(30.6,63.5){(a-2)}
       \put(55,63.5){(a-3)}
       \put(79.4,63.5){(a-4)}
       \put(6.4,42){(b-1)}
       \put(30.6,42){(b-2)}
       \put(55,42){(b-3)}
       \put(79.4,42){(b-4)}
       \put(6.4,21){(c-1)}
       \put(30.6,21){(c-2)}
       \put(55,21){(c-3)}
       \put(79.4,21){(c-4)}
    \end{overpic}
    \caption{\label{fig:Q2} The injected beam charge vs.\ laser focal waist radius $w_0$. The plasma density $n_p$ for (a) (b) (c) equals to $2\times 10^{18}\rm cm^{-3}$, $4\times 10^{18}\rm cm^{-3}$, $6\times 10^{18}\rm cm^{-3}$, respectively. And the laser strength parameter at focus $a_0$ for (x-1 to 4) equals to 12, 10, 8 and 6, respectively, where x stands for a, b or c. This figure shares the same legend as Fig.~\ref{fig:Q1}.
}
\end{figure}

 The charges of the beams injected solely within $z_{fe}<z<z_{ie}$ for difference cases are shown in Figs.~\ref{fig:Q1} and \ref{fig:Q2}. One can see that the charge estimation Eq.~(\ref{eq:Q}) with $C=1.19\ \times10^{-18}\ \rm nC\cdot cm^2$ has a good agreement with the simulations.
 
\begin{figure}
    \centering
    \begin{overpic}[width=0.45\textwidth]{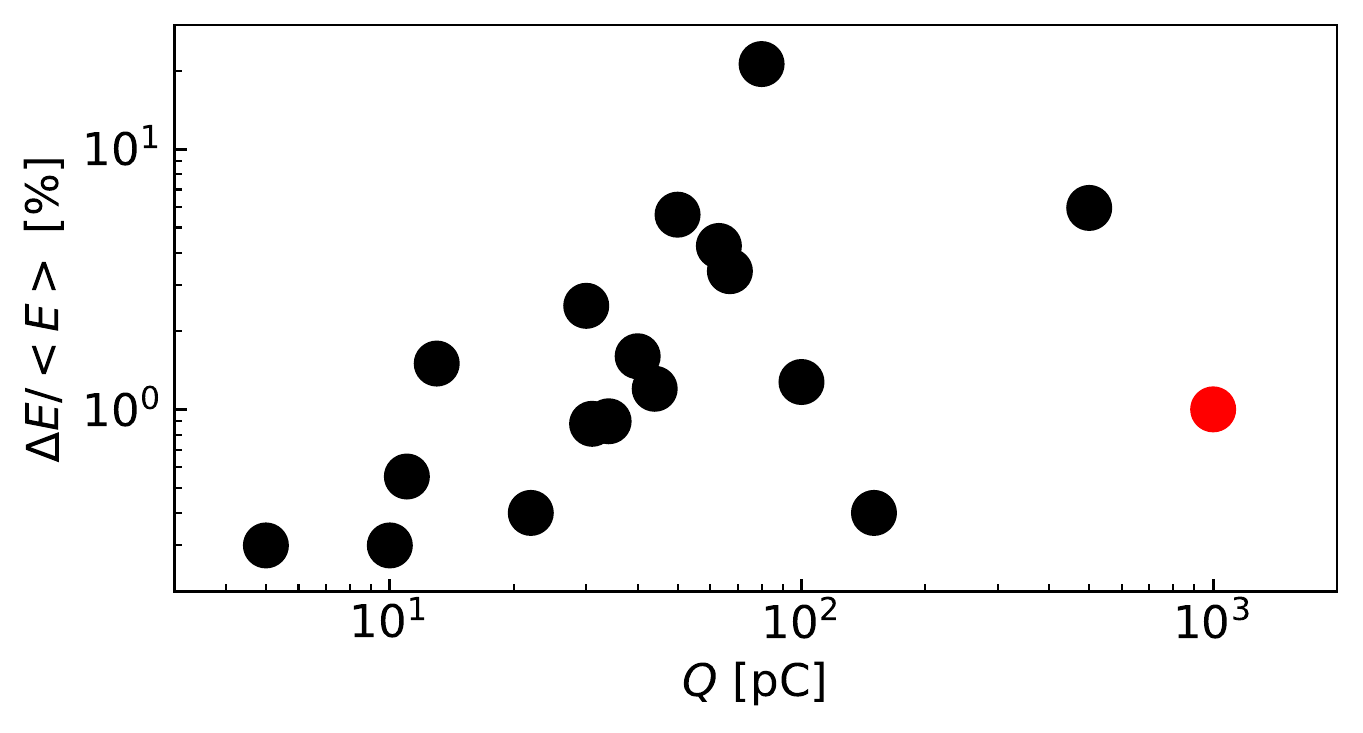} % 38 would be changed after all the refs being confirmed
    \end{overpic}
    \caption{\label{fig:es_vs_q} The energy spread vs.\ charge in previously published papers (black dots) and in this paper (red dot).
}
\end{figure}

\section{\label{sec:level5}conclusion}
In conclusion, we have proposed an improved self-injection scheme in a laser wakefield accelerator with tightly focused drive laser pulse, which produces an electron beam with the charge in the order of nanocoulomb, the energy spread in the order of 1\% and the normalized emittance in the order of $0.1\ \rm mm\cdot mrad$ by a $\sim100\ \rm TW$ laser. The key reason for such optimization is that the injection is induced by the laser defocusing shortly after the focal position, instead of by highly nonlinear laser evolution. We have also found an empirical formula for prediction the charge of the injected beam, which has good agreements with the PIC simulations. This work is a follow-up of our previous studies on the simultaneous optimization of both the energy spread and the charge of the LWFA produced electron beam~\cite{WangJPPCF2022,WangJMRE2022}. As shown in Fig.~\ref{fig:es_vs_q}, this proposed scheme has a significant multi-parameter optimization for the beam energy spread and charge compared with previous results~\cite{LeeNP2006,WangNC2013,KimPRL2013,JalPRL2021,LiPPCF2020,RecPRL2009,JPCnc2017,WWTPRL2016,KeLPRL2021,KirPRL2021,KeLAS2021,ZengPRL2015,ZengNJP2020}. Such optimization, realized by such simple experimental setup, may broaden the application range of LWFAs.

\begin{acknowledgments}
This work is supported by Research Foundation of Institute of High Energy Physics, Chinese Academy of Sciences (Grant No.\ E05153U1, No.\ E15453U2, No.\ E329A1M1, No.\ Y9545160U2 and No.\ Y9291305U2), and Key Research Program of Frontier Sciences of Chinese Academy of Sciences (Grant No. QYZDJ-SSW-SLH004).
\end{acknowledgments}

\bibliography{main}% Produces the bibliography via BibTeX.

\end{document}